\documentclass[unsrtnat]{sn-jnl}%

\jyear{2022}%

\theoremstyle{thmstyleone}%

\theoremstyle{thmstyletwo}%

\theoremstyle{thmstylethree}%

\usepackage{amsmath,amssymb,amsfonts,bbm}
\usepackage{dsfont} %
\usepackage{graphicx}

\usepackage{url}
\usepackage{soul}
\usepackage[nameinlink,capitalize]{cleveref}
\usepackage[font=small]{caption}

\usepackage{algorithm}
\usepackage{array}
\usepackage{indentfirst}
\usepackage{comment}
\usepackage{booktabs} %
\usepackage{etoc} %
\usepackage{multirow}
\usepackage{makecell}
\usepackage{enumitem}
\usepackage[numbers]{natbib}

\usepackage[export]{adjustbox}
\usepackage{subfig}
\usepackage{epstopdf}
\DeclareGraphicsExtensions{.pdf,.png,.eps,.jpg,.tif,.tiff,.ps}
\graphicspath{{img//}{img/fig1-model-schema//}{img/fig2-transfer-learning//}{img/T2-hate-prediction//}{img/T3-map-of-hate//}{img/T4-embeddings-contextualization//}{img/T5-what-makes-it-hateful//}{img/T6-incremental-training//}{img/T8-other//}{.//}}

\usepackage{soul}

\definecolor{navy}{rgb}{0.1, 0.1, 0.8}
\definecolor{gray}{rgb}{0.6, 0.6, 0.6}
\definecolor{myblue}{rgb}{.8, .8, 1}
\definecolor{olive}{rgb}{0.1, 0.5, 0.1}

\newcommand{\eat}[1]{}

\newcommand{\secmoveup}{\vspace{-0.mm}} %
\newcommand{\squishlist}{
	\begin{list}{$\bullet$}
		{ \setlength{\itemsep}{0pt}
			\setlength{\parsep}{3pt}
			\setlength{\topsep}{3pt}
			\setlength{\partopsep}{0pt}
			\setlength{\leftmargin}{1.5em}
			\setlength{\labelwidth}{1em}
			\setlength{\labelsep}{0.5em} } }
	
\newcommand{\squishlisttwo}{
	\begin{list}{$\bullet$}
		{ \setlength{\itemsep}{0pt}
			\setlength{\parsep}{0pt}
			\setlength{\topsep}{0pt}
			\setlength{\partopsep}{0pt}
			\setlength{\leftmargin}{1.5em}
			\setlength{\labelwidth}{1.5em}
			\setlength{\labelsep}{0.5em} } }
	
\newcommand{\squishend}{
\end{list}  }

\usepackage{xspace}
\newcommand{\dnn}{{HateNet}\xspace}
\newcommand{\tdnn}{{t-HateNet}\xspace}
\newcommand{\dav}{{\sc Davidson}\xspace}
\newcommand{\was}{{\sc Waseem}\xspace}

\newcommand{\titlename}{Transfer Learning for Hate Speech Detection in Social Media} %

\begin{document}

\title{\titlename}

\author*[1]{\fnm{Lanqin} \sur{Yuan}}\email{lanqin.yuan@student.uts.edu.au}

\author[2]{\fnm{Tianyu} \sur{Wang}}\email{tianyu.wang2@anu.edu.au}

\author[2]{\fnm{Gabriela} \sur{Ferraro}}\email{Gabriela.Ferraro@anu.edu.au}
\author[2,3]{\fnm{Hanna} \sur{Suominen}}\email{hanna.suominen@anu.edu.au}
\author[1,2]{\fnm{Marian-Andrei} \sur{Rizoiu}}\email{marian-andrei.rizoiu@uts.edu.au}

\affil*[1]{\orgname{University of Technology Sydney},  \city{Sydney}, \state{NSW}, \country{Australia}}
\affil[2]{\orgname{The Australian National University},  \city{Canberra}, \state{ACT}, \country{Australia}}
\affil[3]{\orgname{University of Turku (UTU)}, \city{Turku}, \state{Southwest Finland}, \country{Finland}}

\abstract{%
Today, the internet is an integral part of our daily lives, enabling people to be more connected than ever before. However, this greater connectivity and access to information increase exposure to harmful content such as cyber-bullying and cyber-hatred. Models based on machine learning and natural language offer a way to make online platforms safer by identifying hate speech in web text autonomously. However, the main difficulty is annotating a sufficiently large number of examples to train these models.
This paper uses a transfer learning technique to leverage two independent datasets jointly and builds a single representation of hate speech.
We build an interpretable two-dimensional visualization tool of the constructed hate speech representation --dubbed the Map of Hate-- in which multiple datasets can be projected and comparatively analyzed.
The hateful content is annotated differently across the two datasets (racist and sexist in one dataset, hateful and offensive in another). However, the common representation successfully projects the harmless class of both datasets into the same space and can be used to uncover labeling errors (false positives).
We also show that the joint representation boosts prediction performances when only a limited amount of supervision is available.
These methods and insights hold the potential for safer social media and reduce the need to expose human moderators and annotators to distressing online messaging.
}

\maketitle

{\let\thefootnote\relax\footnote{This version of the article has been accepted for publication, after peer review (when applicable) but is not the Version of Record and does not reflect post-acceptance improvements, or any corrections. The Version of Record is available online at: \url{http://dx.doi.org/10.1007/s42001-023-00224-9}}}

\secmoveup
\section{Introduction}

\noindent Ubiquitous access to the Internet brought with it profound change to our lifestyle:
information and online social interactions are at our fingertips; 
however, it brings new challenges, such as the unprecedented liberalization of hate speech.
Hate speech is a type of online abuse defined as ``public speech that expresses hate or encourages violence towards a person or group based on race, religion, sex, or sexual orientation''~\cite{cambridge:hate-speech},
Hate speech proliferation is suspected to be an important culprit in creating a state of political violence and in exacerbating ethnic violence, such as the Rohingya crises in Myanmar~\cite{reuters-rohingya}.
Hate speech has also been linked to its victims' health deterioration; 
several studies confirmed that racism and sexism are associated with poorer mental health, including depression, isolation, anxiety, and poor self-esteem \cite{Barreto:2009,Paradies:2015}.
Considerable pressure is mounting on social media platforms to timely detect and eliminate hate speech, alongside cyber-bullying and offensive content~\cite{Zhang:2018}.

This work addresses three open questions about detecting hateful content (i.e., hate speech, racist, offensive, or sexist content).
The first question concerns constructing a more generic detection system for textual hate content.
There is a considerable amount of work on detecting hate speech~\cite{Cheng:2015,waseem2016,Chatzakou:2017,Wulczyn:2017,Davidson:2017,fehn-unsvag-gamback:2018}; 
however, many works rely on hand-crafted features, user information, or platform-specific metadata, which limits its generalization to new data sets and data sources.
The first question is \textbf{can we design a \emph{general-purpose} hate embedding and detection system which does not rely on expensive hand-crafted features and can adapt to a particular learning task?}
The second question relates to data availability.
It is difficult to build large-scale datasets of online hate speech, as platforms usually report and remove such content.
While this has the advantage of protecting the users, it also hampers researchers' efforts to build datasets.
When such datasets are built, they are often of small scale and not representative of the entire spectrum of hateful content.
The question is \textbf{can we leverage multiple smaller, unrelated data sets to learn jointly and to transfer information between apparently unrelated learning tasks?}
The third question relates to the interpretation and analysis of hate speech by asking
\textbf{can we construct a tool for separating types of hate speech and characterizing what makes particular language hateful?}

This paper addresses the above three open questions by jointly leveraging two unrelated hate speech data sets.
We address the first two open questions by proposing \tdnn, a novel neural network transfer learning pipeline.
We train the system by predicting, for each dataset, whether a text is hateful or not.
The system contains shared components that construct the textual embeddings and dataset-specific components that predict the final labels.
We use pre-trained word embeddings,
that we adapt to the current learning tasks using a \emph{bidirectional Long Short-Term Memory} (bi-LSTM) \cite{HochreiterandSchmidhube1997} unit.
This creates a single representation space capable of successfully embedding hateful content for multiple learning tasks.
We show that the system can transfer knowledge from one task to another, thus boosting performance.
The system operates solely on the analyzed text, making it platform and source independent.
Given a new learning task, the hateful embeddings trained on other tasks can be directly applied; 
if labeled data is available for the new task, it can be used to contextualize the embeddings.
We address the third open question by building the \emph{Map of Hate}, a two-dimensional representation of the hateful embeddings described above.
We use this embedding for several tasks.
First, we visually show that hate classes are more separable using the joint embeddings constructed by \tdnn than if trained individually.
Second, the \emph{harmless} classes across the two datasets -- containing texts not labeled as hate speech -- overlap in the joint representation.
Finally, we use the Map of Hate to diagnose mislabeled examples and uncover systematic labeling issues.

\textbf{The main contributions of this work are as follows.}
First, we assemble \emph{\dnn}---a deep neural network architecture---capable of creating task-specific word and sentence embeddings without the need for expensive hand-crafted features.
Second, we propose \emph{\tdnn}, which connects the \dnn architecture with \emph{transfer learning} methods that allow leveraging several smaller, unrelated data sets to construct a general-purpose hate speech embedding.
Third, we introduce the \emph{Map of Hate} -- an interpretable 2D visualization of hateful content, capable of separating different types of hateful content and explaining what makes text hateful.

\secmoveup
\section{Background and prerequisites}
\label{BGSec}

Despite its recency as a research field~\cite{Fortuna2018}, online abuse has received considerable attention;
its subproblems range from quantifying bias and stereotypes in detection models~\cite{Badjatiya2019}, to building dedicated web applications for actively reporting hate speech~\cite{Lange2018}, to organizing shared tasks for aggression identification~\cite{Kumar2018}, to visualizing the geographical spread of hate speech~\cite{capozzi2020contro}, and building novel formulations of the hate speech type identification problem using fuzzy ensemble approaches~\cite{Liu2019}.
In our presentation of related works, we concentrate on hate speech detection in the English language (\cref{subsec:hate-speech-detection}), hate speech mapping, visualization and interpretation (\cref{subsec:hate-speech-mapping}), and domain adaptation (\cref{subsec:transfer-learning-related}).

\secmoveup
\subsection{Hate Speech Detection}
\label{subsec:hate-speech-detection}

\noindent\textbf{Feature-based classifiers.}
Early approaches from 2015 to 2017 mainly used hand-crafted features together with simple off-the-shelf classifiers.
For example, Waseem and Hovy \cite{waseem-hovy-2016} used a \emph{Logistic Regression model} with \emph{character level features} to classify tweets --- short messages from Twitter, a major social media platform.
Davidson et al \cite{Davidson:2017} have also used this modeling method (i.e., Logistic Regression) for tweet classification, but with \emph{word level features}, \emph{part-of-speech}, \emph{sentiment}, and some \emph{meta-data} associated with the tweets.
\emph{User features} (e.g., number of friends, followers, gender, geographic location, anonymity status, active vs. non-active status, among others) have also been shown to be useful in identifying aggressive and anti-social behavior \cite{Cheng:2015,waseem2016,Chatzakou:2017,Wulczyn:2017,Pitsilis2018}. 
These feature-based classifiers have several shortcomings.
First, Fehn Unsv{\aa}g and Gamb{\"a}ck \cite{fehn-unsvag-gamback:2018} have shown that user features only slightly improve the classifier's ability to detect hate speech when tested on three Twitter data sets with a Logistic Regression model.
Second, user features are often unavailable or not uniformly available across all social platforms.

\noindent\textbf{Neural network models.}
In recent years, the research community has shifted away from traditional feature-based classification models towards neural network models, which we introduce briefly.
\emph{Recurrant Neural Networks} (RNN) \citep{rnn_citation} are a form of neural network which utilize an internal memory state to allow neural nodes to process input data of a variable length.
\emph{Convolutional Neural Networks} (CNN) \citep{lecun2015deep} are models that rely on convolution, the mathematical operation to process slices of the input data rather than all of the input data at once by dividing the input data based on some notion of locality.
\emph{Long Short-Term Memory} (LSTM) \citep{lstm_citation} networks are an extension of RNNs that address the RNNs' inability to process long sequences by adding gate parameters that control how much impact the input data has on the internal memory state.
\emph{Transformers} \citep{transformers_citation} are a newer model class that uses the self-attention mechanism to replace recurrence and determine which parts of the input data are important for the subsequent layers.

Neural network models do away with the need for manual feature engineering and instead rely on the model itself to learn useful features.
Badjatiya et al \cite{Badjatiya2017} compared three approaches (CNN, LSTM and FastText) for constructing contextualized word embeddings for potentially hateful words.
Park and Fung \cite{Park-fung:2017} have used a neural network approach with two binary classifiers: a \emph{Convolutional Neural Networks} (CNNs) with \emph{word and character-level embeddings} for predicting abusive speech, and a Logistic Regression classifier with \emph{\emph{n}-gram features} for discriminating between different types of abusive speech (i.e., racism, sexism, or both).
Zhang et al \cite{Zhang:2018} have applied pre-trained word embeddings and CNNs with \emph{Gated Recurrent Units} (GRUs) to model long dependencies between features.
Founta et al \cite{Founta:2018} have built two neural classifiers: one for textual content and another one for user features. 
Their experiments concluded that joint training of the networks increases the overall performance. 
Pereira-Kohatsu et al \cite{s19214654} presented a LSTM-MLP based model dubbed \textit{HaterNet} to hate speech in Spanish. Despite similarities in the model name (\textit{HaterNet} vs our model \dnn), our work differs in its use of multiple datasets and transfer learning during training. 
More recently, transformer-based models using attention have become popular due to their strong performance across various natural language tasks. 
These models usually utilize large-scale pretraining on large text corpora and require fine-tuning to adapt the model toward a more specialized domain. 
Mozafari et al \cite{10.1007/978-3-030-36687-2_77} explored fine-tuning methods and successfully fine-tuned a BERT model pretrained on English Wikipedia and BookCorpus for the task of hate speech detection. 
Awal et al \cite{awal2021angrybert} applied multitask learning to a BERT-based model with shared and private parameter layers, achieving strong performances.

\secmoveup
\subsection{Hate Speech Mapping, Visualization and Interpretation}
\label{subsec:hate-speech-mapping}

\noindent \textbf{Interpretation.}
An often-quoted shortcoming of neural network methods is their lack of interpretability, and research has been devoted to interpreting their decision-making and results.
Park and Fung \cite{Park-fung:2017} have \emph{clustered} the vocabulary of the Waseem \cite{waseem2016} data set using the fine-tuned embedding from their model and found the clusters grouped sexist, racist, harassing etc. words.
Wang \cite{Wang:2018} has presented the following three methods for interpretability: 1) \emph{iterative partial occlusion} (i.e., masking input words) to study the network sensibility to the input length; 2) its opposite problem, called \emph{lack of localization}, in which the model is insensitive to any region of the input; and 3) \emph{maximum activations} of the final max-pooling layer of a CNN-GRU network to identify the lexical units that contribute to the classification.
According to their results, long inputs are more challenging to classify, and not all the maximum activated units are hateful.

\noindent\textbf{Visualization.}
There have been several attempts at visualizing the hatefulness of speech with the aid of machine learning classification models. 
Capozzi et al \cite{capozzi2020contro} employed an SVM model on an Italian hate speech Twitter dataset to generate word occurrence and co-occurrence visualizations, as well as Choropleth and Dorling maps to visualize the locations of where hateful tweets originate geographically. 
Modha et al \cite{modha2020} explored several traditional and neural network classifiers for hate speech classification as either ``Non-Aggressive", ``Covertly Aggressive", or ``Overtly Aggressive". 
They created a browser plugin that uses the final output of the models to color text based on the model classification; 
it displays a confidence score for each of the 3 classes and gives the option to automatically hide text that was identified as aggressive. Pereira-Kohatsu et al \cite{s19214654} visualized the term frequency, user frequency, mention network, and term embedding of their LSTM-MLP model using T-SNE. Our work differs in our model architecture and we focus on the visualization of the sentence embeddings over terms.

Our work contributes the Map of Hate, a visualization that allows us to determine the type of speech employed (racist, sexist, etc.) and diagnose labeling errors.

\secmoveup
\subsection{Domain Adaptation}
\label{subsec:transfer-learning-related}

\noindent\textbf{Frustratingly Easy Domain Adaptation.}
Karan and {\v{S}}najder \cite{Karan-snajder:2018} have applied the \emph{frustratingly easy domain adaptation} (FEDA) framework \cite{Daume-iii:2007} to hate speech detection.
This method works by joining two data sets from different domains in which their features are copied three times, depending on their presence in one or both data sets.
The study concluded that domain adaptation boosts the classification performance significantly in six out of the tested nine cases (or data sets).

\noindent\textbf{Multi-task learning.}
Multi-task learning addresses two or more tasks simultaneously while sharing knowledge between the tasks in order to improve performance on each of them. 
The core idea is that learning multiple similar tasks acts as a regularizer for the joint model, which is less prone to over-fitting and more generalizable.
Waseem et al \cite{Waseem2018} use domain adaptation in a multi-task learning framework to jointly learn from different data sets.
They show that this can mitigate overfitting in a hate speech detection context, a common problem when training models with small data sets. 
They do not utilize pretrained embeddings in their work. 
Kapil and Ekbal \cite{KAPIL2020106458} develop four different neural network models trained using a multi-task learning setup. 
They were able to achieve strong classification performance across all their experiments compared to baselines and state-of-the-art models. 
They differ from our work in model architecture as they utilize both shared weights and task-specific parameters. 
Rajamanickam et al \cite{rajamanickam2020joint} developed two separate models: a double encoder based on LSTMs and a hard parameter sharing LSTM model, similar to our work. 
Our work differs from the above as we construct one prediction head per dataset, which improves prediction performances.
While multi-task learning can be successfully applied to detect abusive language, care must be taken during training to make sure that one task does not dominate the other tasks.
Also, it does not produce features and models that can be easily applied to other problems.

\noindent \textbf{Transfer learning} involves using features, weights, or any knowledge acquired for one task to solve another related problem.
Transfer learning has been extensively used for domain adaptation and building models to solve problems where only limited data is available \cite{Pan:2010}.
Formally, transfer learning involves the concepts of domains and learning tasks.
Given a \emph{source domain} $D_S$ and \emph{source learning  task} $T_S$, a \emph{target domain} $D_T$ and \emph{target learning task} $T_T$, transfer learning aims to make a contribution (i.e., improvement) to the learning of the \emph{target predictive function} $f_T(\cdot)$ in $D_T$ using the knowledge in $D_S$ and $T_S$ where $D_S \neq D_T$, or $T_S \neq T_T$ \cite{Pan:2010}.
In this work, the label distribution of the two tasks is different, but related, since the two used data sets are annotated for analyzing different types of hate speech.
$Y_S \neq Y_T$ where $Y$ are the classes to be learned for the two tasks, respectively.

Transfer learning in hate speech detection has been previously explored. 
Pamungkas and Patti \cite{pamungkas-patti-2019-cross} examined the possibility of transfer learning across hate speech datasets of different languages using an LSTM-based model. 
They found that there was some ability for the transfer from other languages to English but models were overall unable to match the performance of a model trained solely on datasets of the target language.
The closest prior work to our own is that of Agrawal and Awekar \cite{Agrawal2018}, who applied transfer for cyberbullying detection --- another form of online abusive behavior.
They experiment with transferring embeddings weights and network weights from models trained in a source domain to models fine-tuned in a target domain.
Their transfer is one-directional (from source to target), which
differs from our own approach (\cref{subsec:transfer-learning}) in which embedding weights are fine-tuned by solving both tasks at once: construct a single representation space and keep the classification tasks specific to each data set.
Using our method, both tasks profit (similar to multi-task learning) and we generate embedding usable for future tasks.

\secmoveup
\section{Model}\label{secModel}
\noindent In this section, we first present our proposed deep neural network architecture, which inputs the raw text of tweets and predicts its hate category (\cref{subsec:model-pipeline}).
Next, we augment the architecture to allow to jointly train and make predictions on multiple data sets (\cref{subsec:transfer-learning}).

\secmoveup
\subsection{The Hate Speech Detection Pipeline}
\label{subsec:model-pipeline}

\begin{figure}[t]
	\centering
	\includegraphics[width=0.95\textwidth]{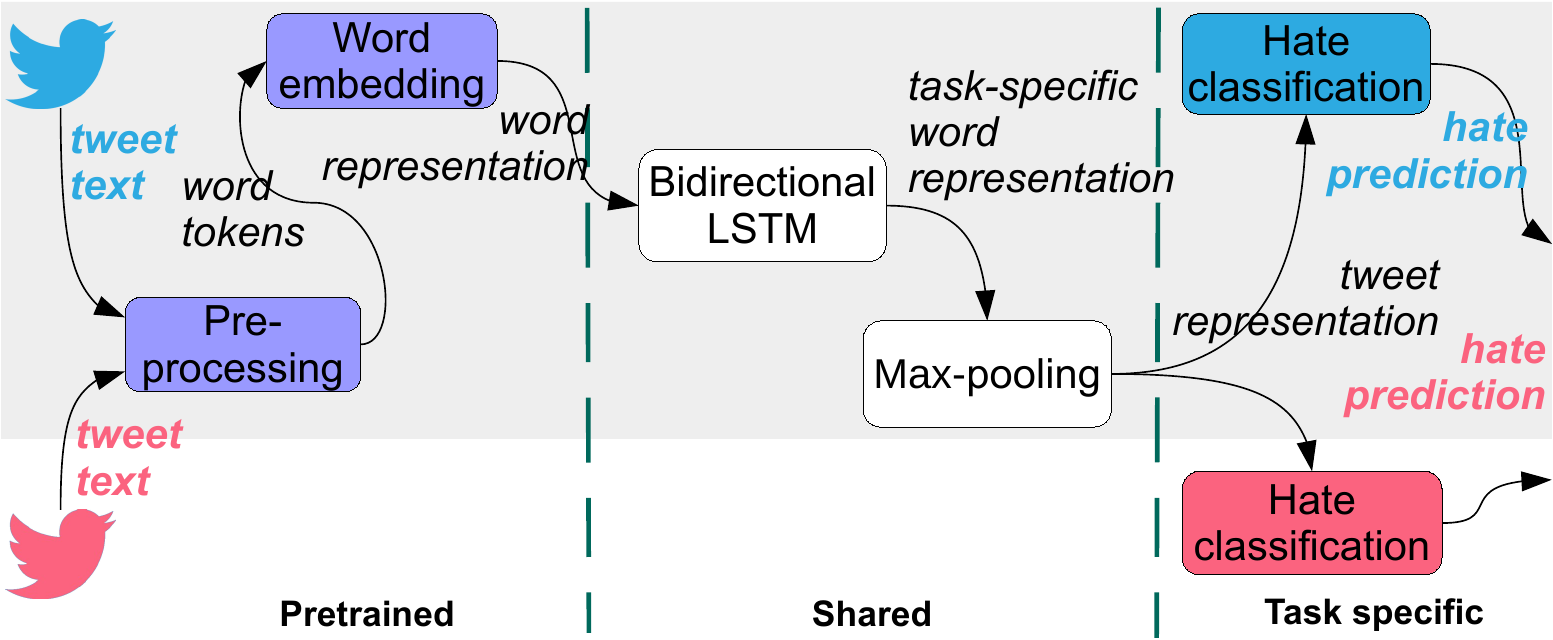}
	
	\caption{
		The conceptual schema of \dnn (on \textcolor{gray}{gray} background) and \tdnn (whole schema) -- our transfer learning architectures.
		\textit{(\dnn)} 
		The text of tweets is processed through two pre-trained units (shown in blue background), and three units whose parameters are trainable end-to-end via back-propagation.		
		The output of the chain is the hate prediction for the input text.
		\textit{(\tdnn)}
		The tweets in two data sets (the \textcolor{red}{red} and the \textcolor{blue}{blue} data set) are processed through a shared pipeline (pre-processing, ELMo, bi-LSTM and max-pooling) and a task-specific component (the Hate classification).
		The \emph{tweet representation} space constructed at the output of the max-pooling unit is adequate for both learning tasks.
	}
	\label{fig:transfer-learning}
\end{figure}

\noindent \cref{fig:transfer-learning} highlighted in gray shows the conceptual schema of our model \emph{\dnn}, which chains the five units detailed below.

\textbf{The pre-processing unit.}
Compared to textual data gathered from other sources, Twitter data sets tend to be noisier. 
They contain misspellings, non-standard abbreviations, Internet-related symbols, slang, and other irregularities.
We pre-process and tokenize the textual data before training our learners.
We remove repetitive punctuation, redundant white spaces, emojis, as well as \emph{Uniform Resource Locators} (URLs).
We add one space before every remaining punctuation.
Note that we do not apply stemming nor remove stopwords.
This results in a cleaner text than the original input.

\textbf{The word embedding unit.}
In order to process natural language text, we map each English term to a numerical vectorial representation --- dubbed the \emph{word representation}.
Training word representations typically requires amounts of textual data larger than our available hate speech data sets.
Therefore, we opt to start from pre-trained word embedding models, among which we select the ELMo~\cite{peters-etal-2018-deep}, as it is less resource-intensive to run compared with alternate technologies such as transformers.
ELMo is itself a \emph{Neural Network} model, which takes as its input a sentence, and outputs a vector representation for each word in the sentence.
Peters et al \cite{peters-etal-2018-deep} show that ELMo's predictive performances are improved when used in conjunction with another (pre-trained) word embedding model.
Here, we use the \emph{Global Vectors} (GloVe) embedding \cite{pennington-etal-2014-glove} pre-trained on a data set of two billion tweets.
ELMo constructs a lookup table between the words observed in the training set and their pre-trained representations.

When encountering a new word not present in the lookup table, ELMo falls back to constructing a character level encoding, starting from its spelling.
Polysemy -- i.e., the fact that a word may have multiple possible meanings -- is another problem for word embedding methods that employ lookup tables, as each word can only have one entry in the table and precisely one representation.
ELMo addresses the issue of possibly multiple (and hence polysemous) vectors by first ``reading" through the whole sentence and then \emph{tweaking} the word representation according to the context; the same word may have different representations in different sentences.
In this work, we employ the pre-trained ELMo 5.5B~\cite{peters-etal-2018-deep} together with the Twitter-trained GloVe embeddings~\cite{pennington-etal-2014-glove}.

\textbf{The bi-LSTM.}
The ELMo model is pre-trained for general purposes, and consequently, its constructed embeddings may have limited usefulness for the hate speech detection application.
Thus, we add a bi-LSTM layer with randomly initialized weights to adapt the ELMo representation to the hate speech detection domain.
The bi-LSTM module scans each sentence twice: from left to right and from right to left.
This scanning produces two task-specific word representations for each word in the sentence -- i.e., one from each scan.

\textbf{The max-pooling unit.}
The next unit in the \dnn pipeline is the max-pooling layer; it constructs the embedding of an entire sentence starting from word representations.
It inputs the two numerical representations of each word from the forward and backward pass of the bi-LSTM; 
it constructs a fixed-length vector by taking the maximum value for each dimension of the word representations.
This results in a sentence representation defined in the same high-dimensional space as the word embeddings.
Based on the output of the max-pooling unit, we can highlight which words contribute most to the classifier decision by counting how many times a word's dimension is selected for the sentence representation.
This can be used as a proxy~\cite{jain-wallace-2019-attention} to isolate the hateful content in a sentence (further discussed in \cref{resultSec}).

\textbf{The hate classification unit.}
The last unit of the pipeline is a differentiable classifier consisting of a fully connected layer with a softmax activation layer.
It inputs the previously constructed sentence representation and outputs the final prediction, which is used to calculate the Cross-Entropy Loss according to the ground truth and train the model.
The weights of all the trainable modules in \cref{fig:transfer-learning} (i.e., the Bi-LSTM, the Max-pooling, and the Hate classifier) are trained end-to-end via back-propagation.
The following section gives the implementation details of each module. 

\secmoveup
\subsection{Transfer Learning Setup}
\label{subsec:transfer-learning}

\noindent Here, we propose \tdnn, an extension of \dnn to predict the type of hate speech in two unrelated data sets. 
The two models share the same base architecture, with the difference being that \tdnn utilizes transfer learning to solve both tasks jointly.
Intuitively, this allows insights learned from one task to be transferred to another, improving performance.

\textbf{A mix of shared and individual processing units.} 
\cref{fig:transfer-learning} shows \emph{\tdnn} the transfer learning schema that we have developed for leveraging multiple data sets and for solving multiple learning problems.
Visibly, \dnn is an integral sub-part of \tdnn.
The pre-processing and the word embedding units are the same as the non-transfer settings, both having pre-trained weights.
The bi-LSTM and the max-pooling units are also shared (i.e., they are processing text from multiple data sets and trainable via back-propagation).
After we obtain the numeric representation for each tweet, we separate the data from different data sets and feed it into identically configured classifiers dedicated to each task.
We make the final predictions for each data set independently from each other, as the prediction classes may be different. 
The entire processing pipeline is shared among all learning tasks apart from the final classification units.
We learn a single word representation and a single sentence representation specific to all the tasks. 
Consequently, we leverage multiple specific data sets to train a larger, more general model for hate speech prediction.

\textbf{Deep learners implementation details.}
Our proposed methods are implemented using the PyTorch library~\cite{paszke2017automatic}. %
We use GloVe with 200 dimensions to initialize the word representations used by ELMo.
The ELMo embeddings have $4,096$ dimensions.
We use a 2-layer stacked bi-LSTM with a hidden vector size of 512 dimensions.
The final fully connected layer has a size of 128.
We minimize the Cross-Entropy Loss during training using the Adam optimizer with the weight decay of $0.001$ and an initial learning rate of $0.001$. 
We train the model for $10$ epochs using a batch size of $350$. 
At every epoch, we test the model using a validation set, and the final weights are selected using the performance on the validation set.
We tune the most critical learning parameters (i.e., the hidden state vector size in the bi-LSTM unit and the batch size of the learning process) via Grid Search on the validation set (see more details in the appendix).

\secmoveup
\section{Datasets and Experimental Setup}
\label{experimentSec}

\textbf{The \was data set} \cite{waseem2016} is publicly available, and consists of $15,216$ instances from Twitter annotated as \textit{Racist}, \textit{Sexist}, or \textit{Harmless}\footnote{Note that both dataset denote the \textit{Harmless} class as \textit{Neither}. 
However, \textit{Neither} has a limited meaning that is contextualized to a given dataset; we show in \cref{resultSec} that the tweets in these classes overlap in our hate representation.
We therefore interpret this class as non-hateful, i.e. \textit{Harmless}.}.
This set is very imbalanced, with the majority class being \textit{Harmeless}, which is meant to reflect a real-world scenario where hate speech is less frequent than neutral tweets.
The data selection appears biased, with most sexist tweets about a single television show in Australia and most racist tweets being Islamophobic.
There are some questions about the quality of labeling in this data set, namely the number of false positives, considering that the data set was compiled to quantify the agreement between expert and amateur raters.
Waseem~\cite{waseem2016} acknowledges this and observes that ``the main cause of the error are false positives''.
Here are some examples of such \textit{sexism} false positives:
{ \small 
\begin{itemize}
   \item 
\begin{verbatim}
@FarOutAkhtar How can I promote gender equality without sounding 
preachy or being a feminazi''? #AskFarhan \end{verbatim}
    
   \item \begin{verbatim}
i got called a feminazi today, it's been a good day. \end{verbatim}
   
   \item 
\begin{verbatim}
Yes except the study @Liberal_fem (the Artist Formerly known as 
Mich_something) offered's author says it does NOT prove bias 
@TamedInsanity \end{verbatim}
   
   \item 
\begin{verbatim}
In light of the monster derailment that is #BlameOneNotAll here are 
some mood capturing pics for my feminist pals \end{verbatim}
\end{itemize}

\textbf{The \dav data set} \cite{Davidson:2017} is also publicly available.
It consists of $22,304$ instances from Twitter annotated as \textit{Hate}, \textit{Offensive}, and \textit{Harmless}.
This data set was compiled by searching for tweets using the lexicon from \href{https://hatebase.org/}{Hatebase.org}.

\begin{comment}
\begin{table}[t]
    \centering
    \caption{Data for hate speech detection on Twitter in English}
    \begin{tabular}{lll}
    \toprule
    Data set & Classes & \#instances \\ \midrule
    \was & \textit{Racist}, \textit{Sexist}, \textit{Harmless} & $15,216$ \\
         
    \dav & \textit{Hateful}, \textit{Offensive}, \textit{Harmless} & $22,304$ \\ \bottomrule
    \end{tabular}
    \label{tab:datasets}
\end{table}
\end{comment}

\begin{figure*}[t]
	\centering
	\newcommand\myheight{0.25} %
	\subfloat[]{
		\includegraphics[height=\myheight\textheight,page=1]{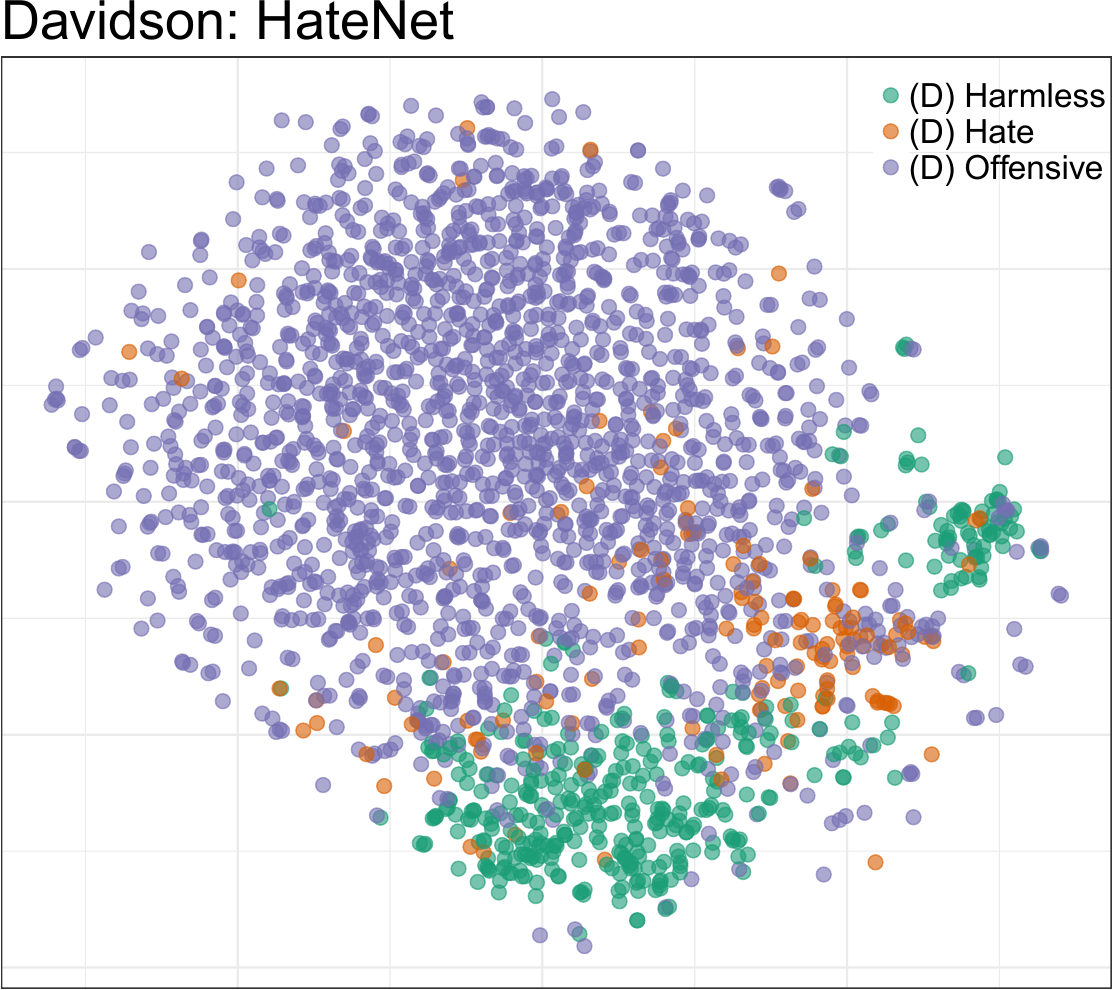}%
		\label{subfig:MOH-dav-dnn}%
	}%
	\subfloat[]{
		\includegraphics[height=\myheight\textheight,page=2]{Davidson_MOH_R}%
		\label{subfig:MOH-dav-tdnn}%
	}\\%
	\subfloat[]{
		\includegraphics[height=\myheight\textheight,page=1]{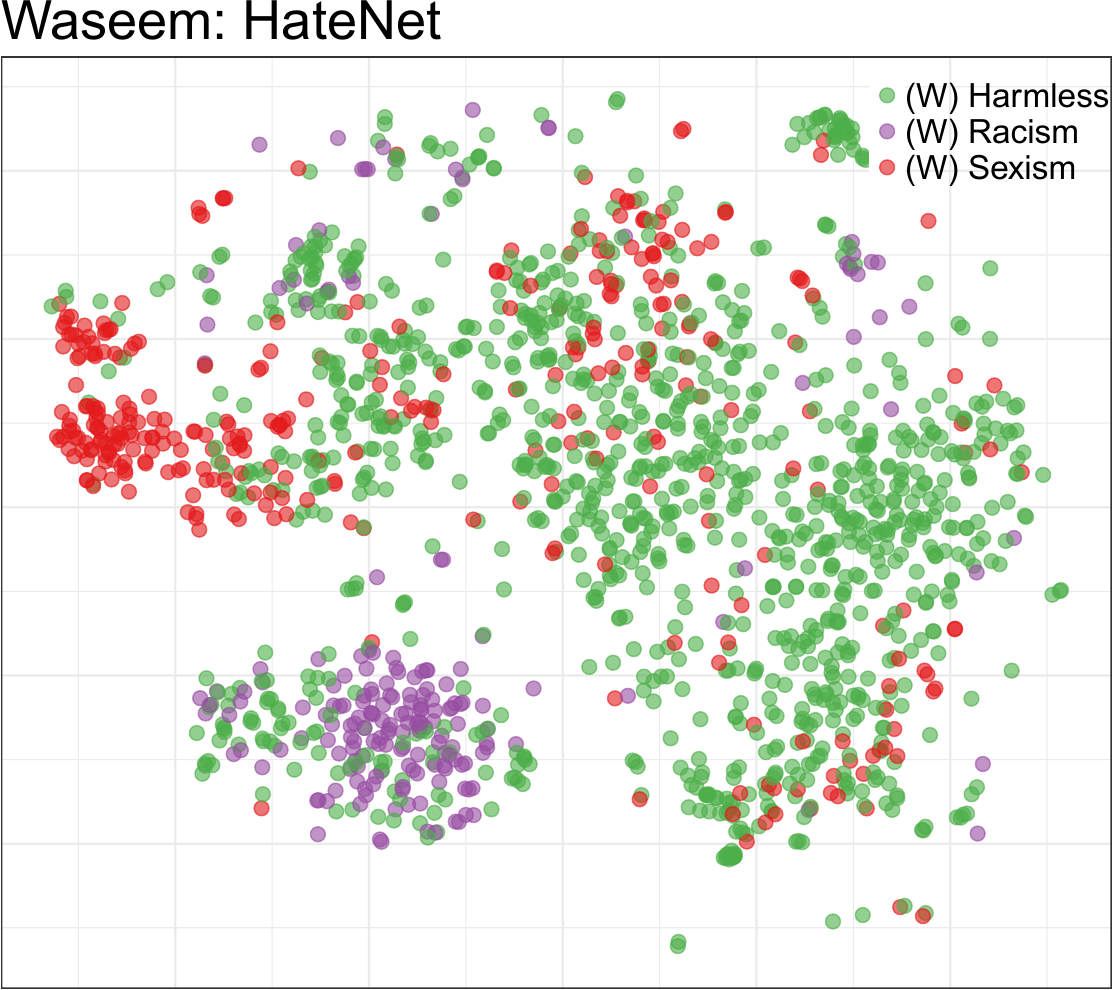}%
		\label{subfig:MOH-was-dnn}%
	}%
	\subfloat[]{
		\includegraphics[height=\myheight\textheight,page=2]{Waseem_MOH_R}%
		\label{subfig:MOH-was-tdnn}%
	}%
	\caption{
      	\textbf{The \emph{Map of Hate}} constructed on the \dav data set (a)(b) and the \was data set (c)(d), by using the tweet embeddings generated by \dnn (a)(c) and \tdnn (b)(d).
		Note that the axes of the Map of Hate are synthetic and not interpretable.
	}
	\label{fig:T3-MOH}
\end{figure*} 
\textbf{Hate speech prediction setup.}
We predict the types of hate speech (\textit{hate}, \textit{offensive} and \textit{harmless} for \dav, and \textit{racism}, \textit{sexist} and \textit{harmless} for \was) using four classifiers: the Davidson and the Waseem baselines, and our approaches \dnn and \tdnn.
Given the class imbalance in both data sets, we over-sample the smaller classes to obtain a balanced data set.
We have tried under-sampling the larger classes, but we obtained worse results.
Unless otherwise stated, we train each classifier on $80\%$ of each data set.
We use $10\%$ for validation and the remainder of $10\%$ for testing.
However, when studying the prediction with minimal amounts of data, we train on as little as $10\%$ of the data set (see \cref{resultSec}).
We repeat the training and testing 10 times, reporting the mean and standard deviation.
\tdnn is trained on $80\%$ of both data sets simultaneously; the other baselines are trained on each data set individually.
We evaluate the prediction performance using the F1 measure -- i.e., the geometric mean of precision and recall.
A classifier must simultaneously obtain high precision and recall to achieve a high F1.
We use the \emph{macro-F1} so that smaller classes are equally represented in the final score.

\textbf{Baselines.}
We compare our proposed method with two baselines: 
\citet{waseem-hovy-2016} and \citet{Davidson:2017}.
\citet{waseem-hovy-2016} pre-processes the text by removing punctuation, excess white spaces, URLs, and stopwords and applying lower-casing and Porter stemmer for removing morphological and inflexional endings from words in English.
They use a Logistic Regression model with character \textit{n}-grams of lengths up to four as features.
\citet{Davidson:2017} applies the same pre-processing as \citet{waseem-hovy-2016} and also uses a Logistic Regression model.
They included several word level features as 1-3 word \textit{n}-grams
weighted with TF-IDF and 1-3 Part-of-Speech \textit{n}-grams.
They also use two types of tweet-level features.
The first type is readability scores, taken from a modified version of Flesch-Kincaid Grade Level and Flesch Reading Ease scores (with the number of sentences fixed to one).
The second type is sentiment scores derived from the VADER sentiment lexicon design for social media \citep{Hutto:2014}.
In addition, they include binary and count indicator features for hashtags, mentions, retweets, URLs, and the number of words, characters, and syllables in each tweet.
As they did not carry out an ablation study, it is unclear which features are most predictive for hate speech.

\begin{figure*}[t]
	\centering
	\newcommand\myheight{0.25}
	\subfloat[]{
		\includegraphics[height=\myheight\textheight]{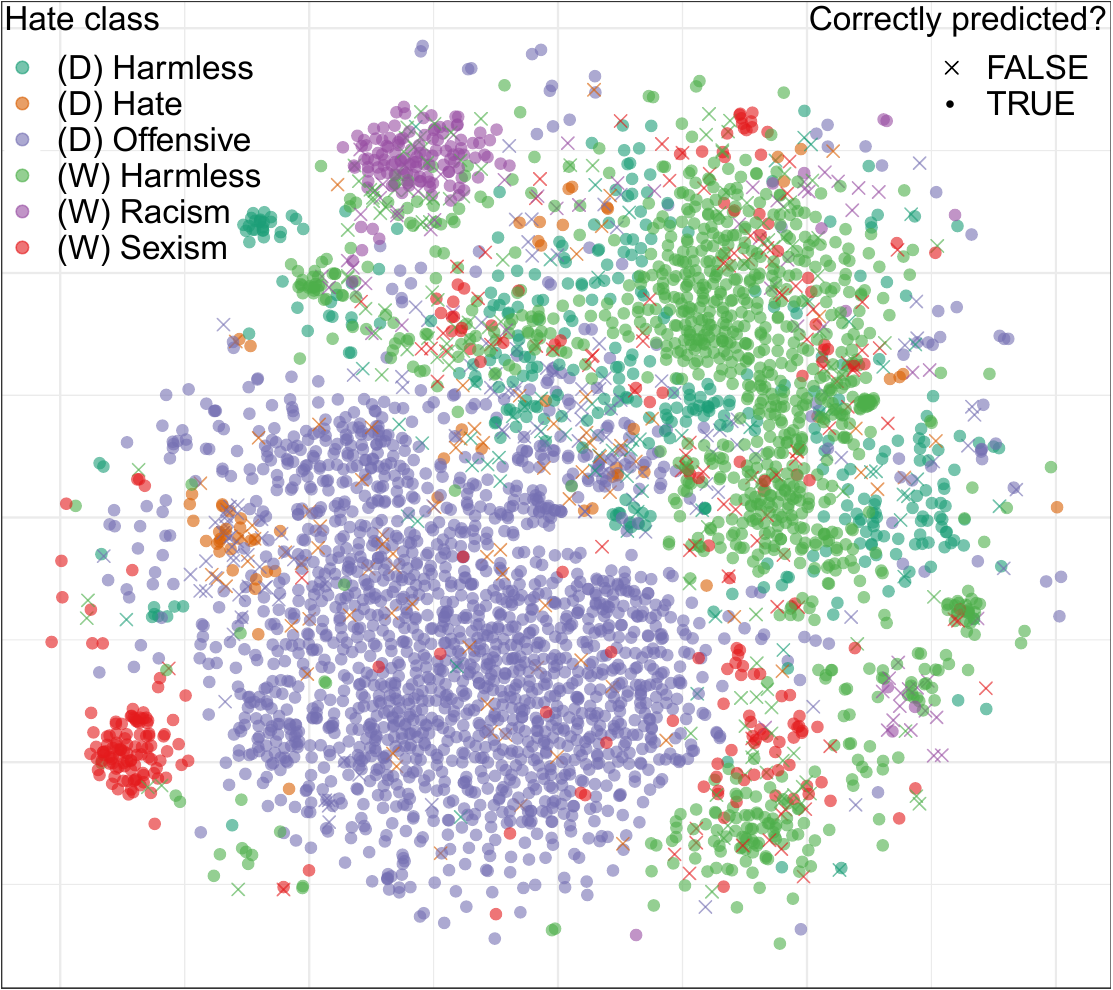}%
		\label{subfig:MOH-dnn-all}%
	}\\%
	\subfloat[]{
		\includegraphics[height=\myheight\textheight]{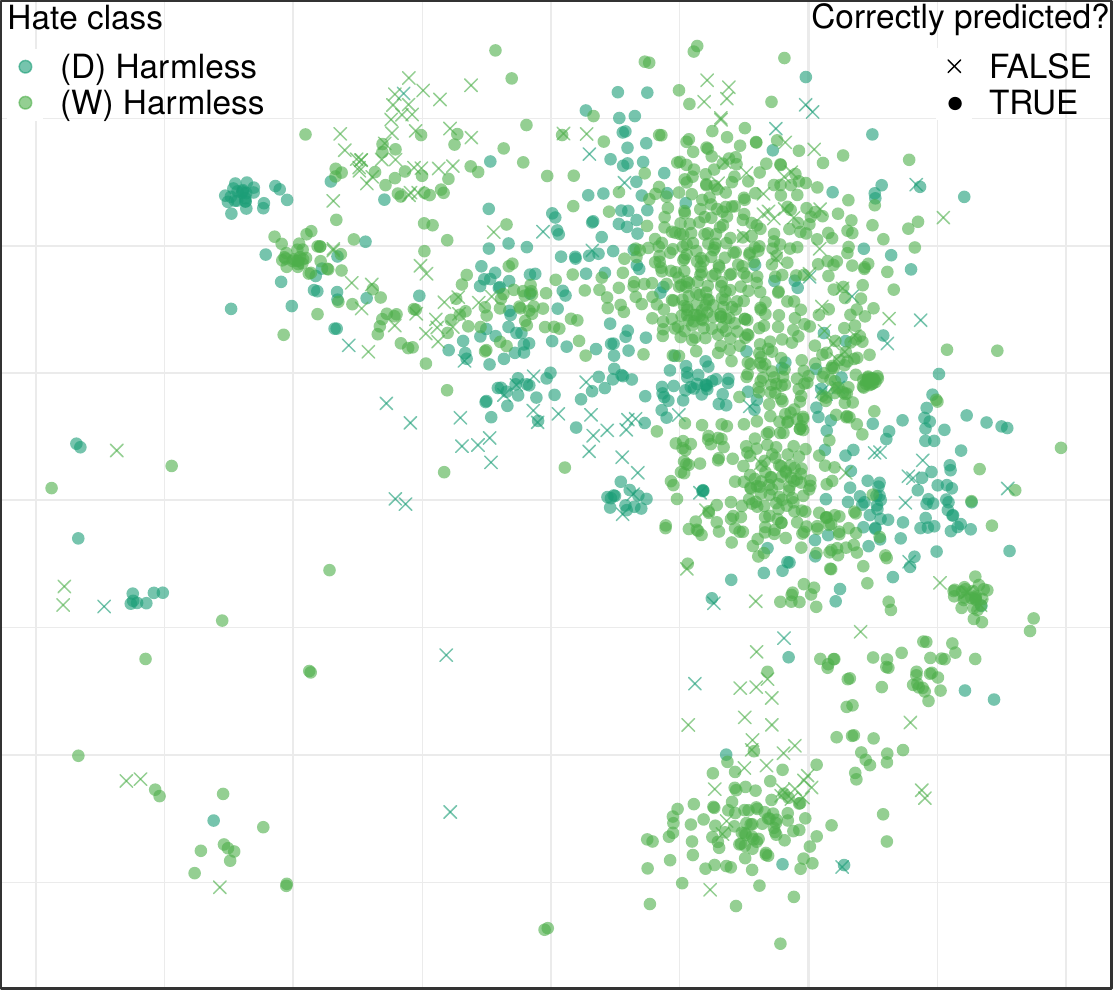}%
		\label{subfig:MOH-tdnn-harmless}%
	}%
	\subfloat[]{
		\includegraphics[height=\myheight\textheight]{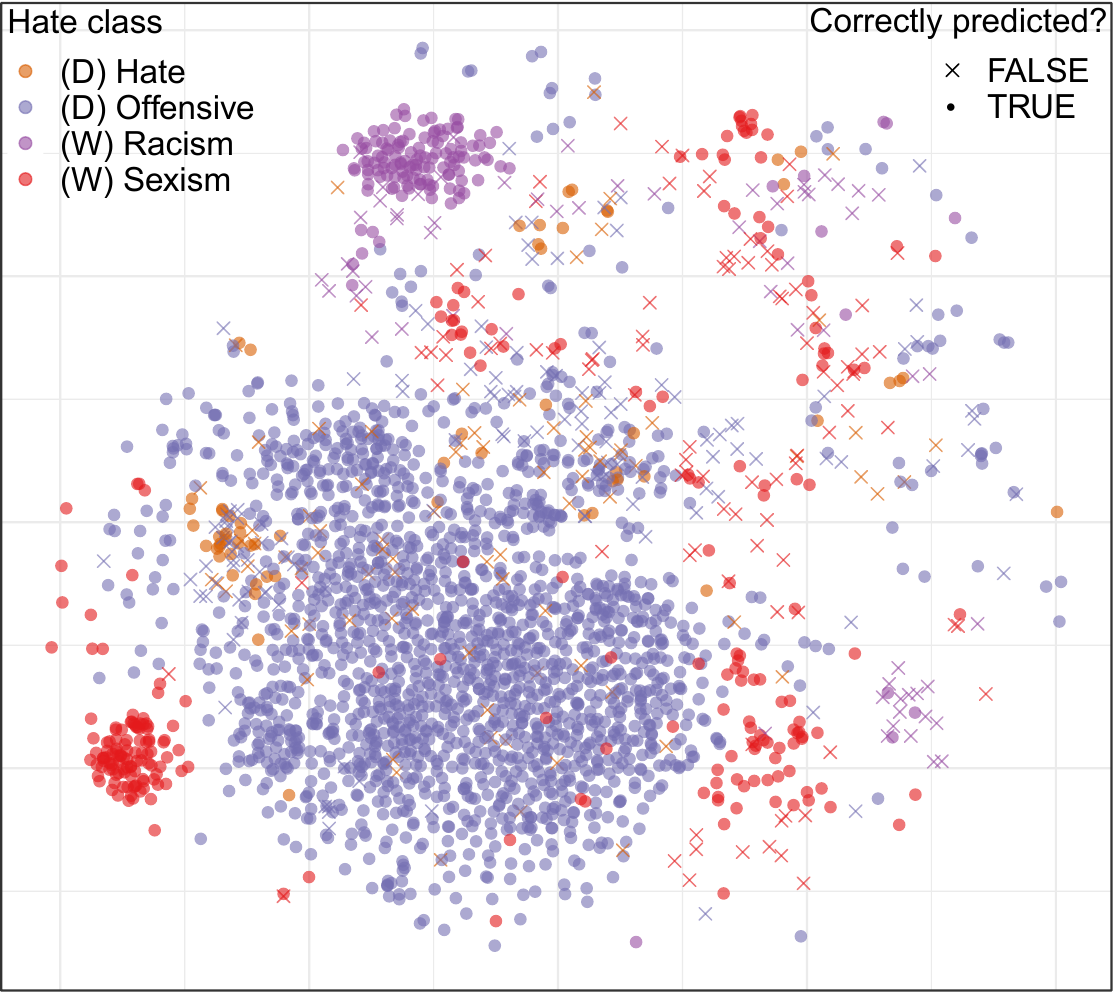}%
		\label{subfig:MOH-tdnn-hateful}%
	}%
	\caption{
      	\textbf{The \emph{Map of Hate} constructed by \tdnn jointly on $10\%$ of the \dav and \was data sets.} 
      	Crosses and circles show incorrectly and correctly predicted examples, respectively. 
      	(interactive version available online at \url{https://bit.ly/39OVBgX}).
      	\textbf{(a)} All the six classes in the two data sets.
      	\textbf{(b)} The Harmless classes of the two data sets overlap.
      	\textbf{(c)} The hateful classes:  Hate and  Offensive (\dav), Racism and Sexism (\was).
	}
	\label{fig:T3-tdnn}
\end{figure*} %

\section{Main findings}\label{resultSec}

\textbf{Map of Hate Visualization.}
To understand the impact of the different modeling choices, we construct the \emph{Map of Hate} -- a two-dimensional visualization of the space of hateful text built using t-SNE~\cite{maaten2008visualizing}, a technique originally designed to visualize high-dimensional data.
Given the tweet representation built by \dnn and \tdnn (the output of the max-pooling units in \cref{fig:transfer-learning}), t-SNE builds a mapping to a 2D space in which the Euclidean pairwise distances correspond to the distance between pairs of tweet representation in the high-dimensional space.
\cref{subfig:MOH-dav-dnn,subfig:MOH-was-dnn} visualize the Map of Hate constructed by \dnn on a sample of $10\%$ of the tweets \dav and \was data sets, respectively.
We observe that the tweets of different classes in \dav appear clustered more closely together than in \was.
Noticeably, the \textit{racist} and \textit{sexist} tweets in \was appear scattered throughout the \textit{harmless} tweets.
We posit this to be indicative of the false positives mentioned by Waseem~\cite{waseem2016}.

The Map of Hate has several usages.
First, it quickly maps each type of hateful content into two-dimensional space regions and explains why a tweet is predicted as hateful. 
This can allow us to uncover patterns in the dataset, such as particular language and labeling errors.
For example, all the tweets in the red bottom-left cluster in \cref{subfig:MOH-was-tdnn} start with ``I'm not sexist, but...'' (many wrongly labeled as sexist, see next).
Second, using the interactive version of the map\footnote{Interactive Map of Hate available online: \url{https://bit.ly/39OVBgX}}
we can help distinguish between the failed and successful cases -- there is a large number of failed predictions in the sexist (red) tweets (\cref{subfig:MOH-tdnn-hateful}) sprinkled in the top-right area of the harmless tweets (observed in \cref{subfig:MOH-tdnn-harmless}).
Interestingly, this might be linked to implicit vs. explicit gender biases;
we know that people with implicit gender bias tend to use phrases like ``I am not sexist'' to begin their sentences \cite{Merritt2010}.
Finally, it can visualize the impact of constructing task-specific tweet embeddings (see the appendix). 

\begin{figure*}[t]
	\centering
	\newcommand\myheight{0.155} %
	\subfloat[]{
		\includegraphics[height=\myheight\textheight]{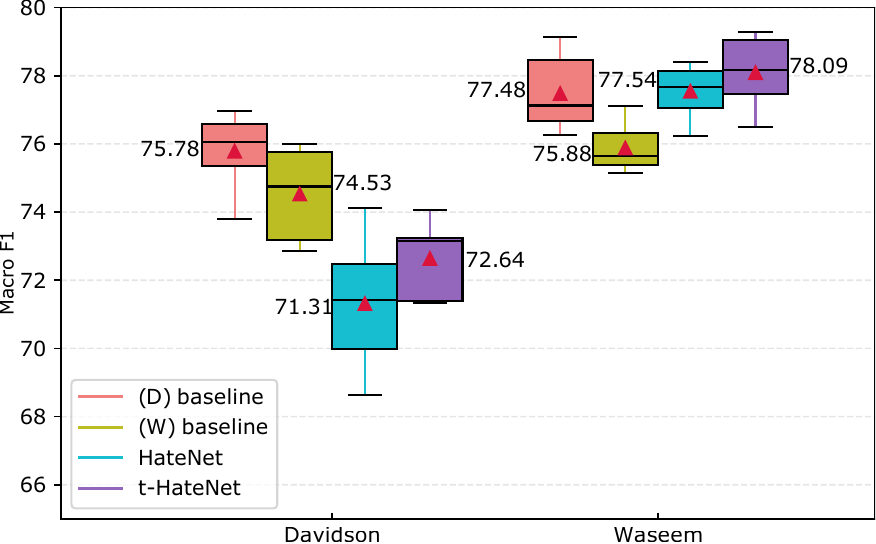} %
		\label{subfig:T2-boxplots}%
	}
	\subfloat[]{
		\includegraphics[height=\myheight\textheight]{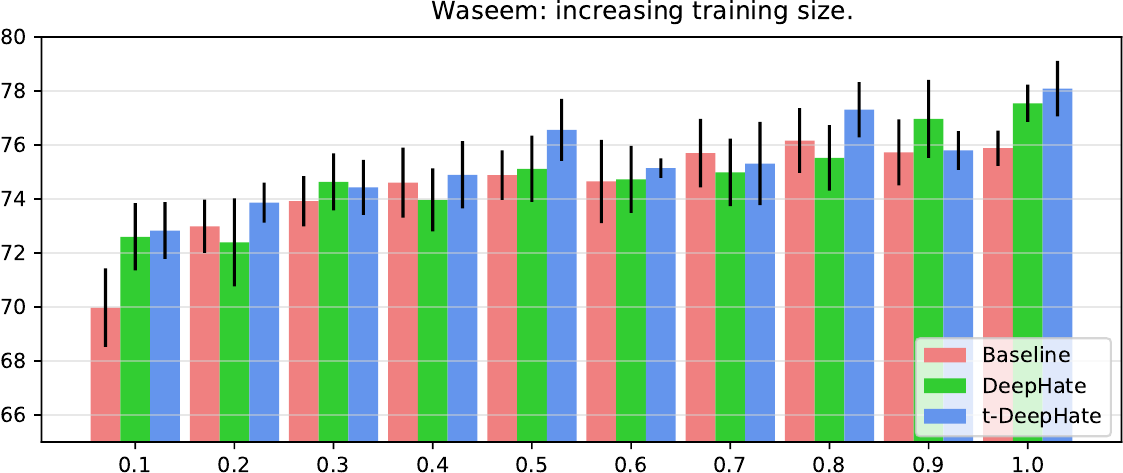}%
		\label{subfig:T6-barplots-waseem}%
	}
	\caption{
		\textbf{(a)} \textbf{Prediction performances on two data sets: \dav and \was.}
		Boxplot summarizing macro-F1 score for each data set, and each approach (baselines, \dnn, \tdnn).
		Red diamonds and values indicate mean F1. Each box consists of 10 independent runs.
		\textbf{(b)} \textbf{Prediction performances with limited amounts of training data} on \was datasets.
		The x-axis show the percentage of the training set used for training, the y-axis shows the macro-F1 measure.
		Each bar shows the mean value over 10 runs, and the standard deviation.
	}
	\label{fig:T2}
\end{figure*} 
\textbf{The effects of transfer learning.}
\cref{subfig:MOH-dav-tdnn,subfig:MOH-was-tdnn} show the tweet sample in \dav and \was data sets, respectively, projected in the joint space constructed by \tdnn.
The \dav tweets appear to lose their clustering and break down into subgroups, especially for the harmless class.
We can interpret this lack of clustering as the class is less distinct in the space, which causes difficulties for the classification head to distinguish them from other classes.
This explains the slightly lower prediction results obtained by \tdnn on \dav.
The situation is reversed on \was, where tweets belonging to different hate categories are clustered more tightly together -- explaining the better performances on this data set.
The decrease in clustering indicates that the quality of representation on \dav is slightly negatively impacted by labeling quality in \was.
However, the prediction on \was benefits from a large increase -- the new space is more adequate to represent its tweets thanks to the \dav data set.

\cref{subfig:MOH-dnn-all} visualizes the tweets from both \dav and \was projected into the \tdnn space in the same figure.
The circles mark correctly predicted tweets by \tdnn while the crosses mark the incorrectly predicted tweets (best seen in the interactive version).
\cref{subfig:MOH-tdnn-harmless} further details only the \textit{harmless} tweets from both data sets, and \cref{subfig:MOH-tdnn-hateful} the hateful tweets from both data sets (\textit{racism} and \textit{sexism} from \was, and \textit{hate} and \textit{offensive} from \dav).
Several conclusions emerge.
Firstly, the hateful and harmless content appears separated in the joint space (bottom-left for hateful and top-right for harmless).
Secondly, the harmless tweets from both data sets appear overlapped (\cref{subfig:MOH-tdnn-harmless}), which is correct and intuitive since both classes stand for the same type of content.
Thirdly, the hateful content (\cref{subfig:MOH-tdnn-hateful}) has a more complex dynamic: the \textit{offensive} (\dav) tweets occupy most of the space, while most of the \textit{sexist} and \textit{racist} (\was) appear tightly clustered on the sides.
This is indicative of the sampling bias in \was.
Lastly, the \textit{sexist} and \textit{racist} (\was) sprinkled tweets throughout the \textit{harmless} are appear overwhelmingly miss-classified, which can be explained by the false positives in \was.

\begin{figure*}[tbp]
	\centering
	\includegraphics[width=1.05\textwidth]{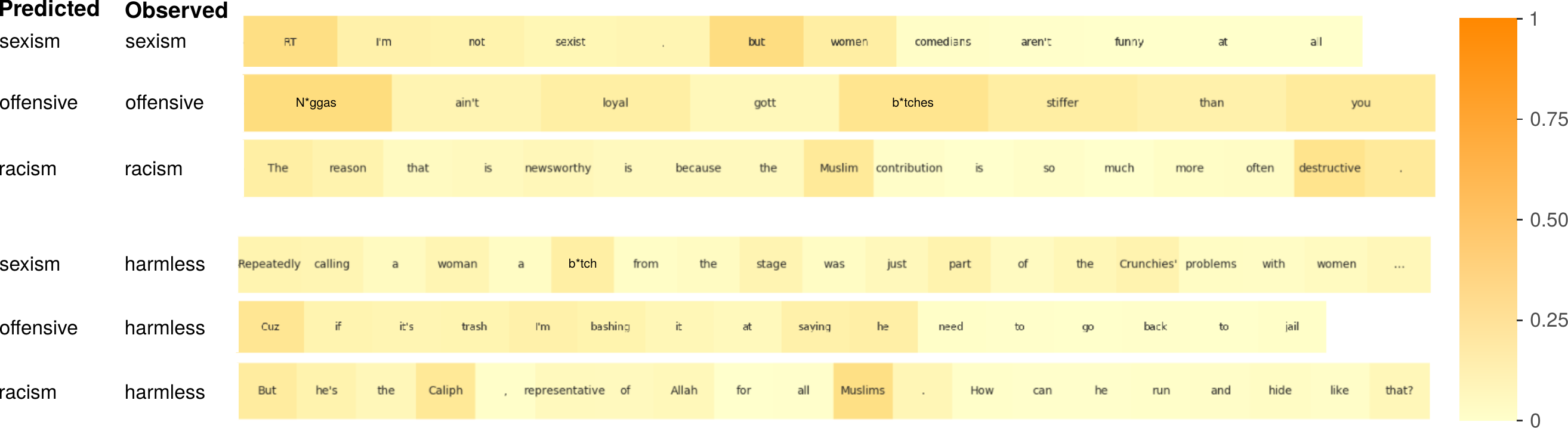}

	\caption{ 
		\emph{Warning: this figure contains real-world examples of offensive language!}
		\textbf{Automatically highlighting offensive terms}.
		It is not possible to illustrate the performance of the technique without quoting actual examples of speech classified as hate speech and to that end we include this figure.
		Examples of six tweets from the \dav and \was data sets, together with their predicted and observed hate category.
		The top three tweets are hateful (\textit{sexist}, \textit{offensive} and \textit{racist} respectively) and their category was correctly predicted.
		The bottom three three tweets are \textit{harmless}, but they were incorrectly predicted as hateful.
		The color map shows how many times a word's representation was selected for the tweet representation, normalized by the size of the embedding (here 512).
	}
	\label{fig:what-makes-it-hateful}
\end{figure*} 
\textbf{Highlighting hateful content.}
Here we describe a method to highlight hateful content in a text.
The max-pooling is a dimension-wise max operation, i.e., for each dimension, it selects the maximum value over all word embeddings in the forward and backward pass of the Bi-LSTM.
Assume a total of $n$ words in the sentence, each represented by a numerical vector with $d$ dimensions.
For each of the $d$ dimensions, the max-pooling picks the maximum value across the $n$ words.
We count the number of times each word is selected to represent the whole sentence across the $d$ dimensions, and we normalize the scores by $d$ (a word can be picked a maximum of $d$ times).
This constructs a score between zero (a word is never selected) and 1 (the word is always selected).
The higher the word's score, the more representative the word's representation for the final sentence representation and the hate prediction.
Despite the recent critiques about linking attention mechanisms to word importance~\cite{jain-wallace-2019-attention}, we find that this method succeeds in highlighting hateful content and can be used to identify patterns in hatefully labeled content, as shown next.

\textbf{Correctly and incorrectly predicted examples.}
\cref{fig:what-makes-it-hateful} shows six examples of tweets in our data sets, with each word highlighted according to its score.  (\emph{Warning! this figure contains real-world examples of offensive language}).
The top three examples are correctly predicted as \textit{sexist}, \textit{offensive} and \textit{racist}, respectively.
We observe that most tweets labeled as \textit{sexism} (\was data set) start with ``I'm not a sexist, but ...'' and variations.
While this might raise questions about data sampling bias in the data set construction, this behavior is captured in the higher weights assigned 
with ``but`` (ending of ``I'm not a sexist, but''), ``woman'' and (strangely) ``RT'' (i.e., retweet).
In the \textit{offensive} example (\dav data set), we notice that offensive slang words are correctly scored higher.
We also notice that \textit{racism} examples (such as the third tweet in \cref{fig:what-makes-it-hateful}) tend to refer exclusively to the Islamic religion and its followers -- which can bias the learned embedding.
However, for this example, we observe that words such as ``destructive'' are correctly recognized as indicators of hate speech.
The bottom three lines in \cref{fig:what-makes-it-hateful} show examples where \textit{harmless} tweets are incorrectly labeled as hateful.
This happens mainly due to their writing style and choice of words.
The tweet misclassified as \textit{sexism} uses language similar to sexism to draw awareness against it, and as a result, it is classified as sexist.
Similarly, the incorrectly labeled \textit{offensive} example is written in a style similar to other offensive tweets in our data set.
Finally, the falsely \textit{racist} tweet uses Islam-related terminology, and because of the data sampling bias, it is classified as racist.

\textbf{Predicting hate speech in two data sets.}
We measure the prediction performances of \dnn and \tdnn against the two baselines, making several observations based on the prediction performances for each classifier and each data set (\cref{subfig:T2-boxplots}).
First, our models \dnn and \tdnn outperform the baselines on the \was data set, and they under-perform them on the \dav data set.
We posit this is due to the baselines' external information -- statistics, user information, and tweet metadata. 
Upon investigating the weights associated with each feature by the logistic regression, we observe that among the 161 features with a non-negligible absolute weight ($>10^{-5}$), 21 features are non-textual and Twitter-related.
The tweet sentiment (quantified using VADER~\cite{Hutto:2014}) is the most relevant, with a high negative weight for \textit{hate} and \textit{offensive} classes, and with a high positive class for the \textit{harmless} class.
Important tweet features include the number of words, user mentions, hashtags, characters, and syllables of a tweet.
We chose not to use this additional information in our approaches, as it would render the obtained results applicable solely to the Twitter data source.
Furthermore, \tdnn confuses the \textit{Hate} class with \textit{Offensive} in 41\% of the cases (more details are included in the appendix).
By manually inspecting some failed predictions, we notice that it is challenging even for humans to differentiate between purposely hateful language (with a particular target in mind) and generally offensive texts (without a target).

Second, we observe that the Davidson baseline outperforms the Waseem baseline on both \dav and \was data sets.
This shows that the external features built by Davidson are more informative for Twitter-originating hate speech than Waseem's.
Third, we observe that \tdnn outperforms \dnn on both data sets, admittedly not by a large margin.

\textbf{Predicting hate speech using limited amounts of data.}
The advantage of jointly leveraging multiple data sets emerges when only limited amounts of labeled data are available.
Here, we study the typical situation in which it is required to learn a hate speech classifier with minimal amounts of labeled data but with the help of a larger unrelated hate speech data set.
We restrain the amount of training data: after sampling the training set ($80\%$ of the data set), the validation set ($10\%$), and the testing set ($10\%$), we further subsample the initial training set so that only a percentage is available for the model training.
For \tdnn, we perform this additional subsampling for only one of the data sets, keeping all training examples of the other data set.
We vary the percentage of training data in the downsampled data set, and \cref{subfig:T6-barplots-waseem} shows the mean prediction performance and its standard deviation on the \was data set (\dav is shown in the appendix).
The performances of \dnn and \tdnn are positively correlated with the size of the training set. 
Visibly, we can observe that \tdnn generally outperforms \dnn across all but two of the ten experiments, showcasing transfer learning's ability to alleviate the issue of limited datasets and improve classification performance.

\secmoveup
\section{Conclusion}\label{conclusionSec}

\noindent With our social interactions and information being increasingly online, more and more emphasis is placed on identifying and resolving Internet issues in our society. 
It is important to make social media safer by detecting and reducing hateful, offensive, or otherwise unwanted social interactions.

In this paper, we introduced \dnn and \tdnn, two machine learning and natural language processing pipelines for differentiating harmless tweets from racist, sexist, hateful, or offensive messages on Twitter. 
\dnn and \tdnn share the same architecture (therefore, the name similarity); 
\tdnn leverages a transfer learning procedure to transfer knowledge from one task to another by constructing a shared generalized representation of hate from both datasets. 
We empirically show that this improves classifier detection and helps to address data scarcity when there is a limited amount of available data. 

We further constructed the Map of Hate, a two-dimensional visualization of the generalized embedding space of hate. We showcase several use cases for the Map of Hate in identifying patterns in hateful speech, identifying errors in labeling in the datasets, identifying cases where the model struggles in its prediction, and in examining the similarity of the datasets through examining the separation of classes between the two datasets in the embedding space.

Our methods contribute to analyzing social media contents at scale to make these web platforms safer and better understand the genre of hate speech and its sub-genres.
Our automated text processing and visualization methods can separate different types of hate speech and explain what makes text harmful. 
Their use could even reduce the need to expose human moderators and annotators to
distressing messaging on social media platforms.

\paragraph{Limitations and future works.}

Here we discuss the two main limitations of this work, their likely causes and possible future steps to alleviate them.

\textbf{Limited performance gain.}
The first limitation relates to our transfer learning architecture's somewhat modest performance improvement, as showcased by our experiments. 
We note, however, that the improvement is consistent across our experiments.
We argue this is linked to the amount of information available to the model -- the two datasets not being fully transferable due to differences in the labeling criteria.
There are two main approaches for increasing the quantity of information available for learning. 
The traditional Machine Learning approach requires increasing the number of labeled examples (i.e., the number of rows in the training dataset). 
This requires significant manual effort and reinforces bias and subjectivity (see next paragraph).
Our work proposes an alternative: increasing the number of smaller, unrelated data sets to learn jointly and to transfer information between apparently unrelated learning tasks.
This provides the required additional information and reduces the overall bias. 
The increase in the number of datasets can be relatively easily achieved as various resources provide hate speech datasets to be added to our framework\footnote{The website \url{https://hatespeechdata.com/} offers a collection of hate speech datasets from multiple works, making them highly accessible for integration.}.

\textbf{Learning human-annotated biases.}
The second limitation concerns Machine Learning models' reliance on bias-rich, human-annotated data.
This limitation applies to all supervised machine learning classification algorithms and is outside the scope of this work.
It is worth noting and discussing some of its consequences for its downstream applications in social science.
We start from the observations that hate speech does not have a universally agreed upon formal definition~\citep{Towards_a_Definition_of_Hate_Speech}.
Furthermore, what individuals perceive as hateful is context-, political-environment, and individual-life-course-dependent.
As a result, human-generated labels of hate speech are subjective and embed biases and preconceptions.
As machine learning models (like the ones we build in this work) learn from examples, they will also learn these inherent biases and further reproduce them in automatic labeling.
Understanding and ideally limiting biases is essential for social science applications before deploying machine-labeled data in downstream analysis.
Without a precise and operationalizable definition of hate speech, we may be unable to circumvent human annotations for model training.

However, our proposed approach partially alleviates the problem.
Learning using multiple datasets can significantly reduce bias and subjectivity in hate speech detection models.
When we increase the size of the training set by coding more data, as is typical in Machine Learning (see previous paragraph), it is crucial to consider the limitations of relying on the same limited set of labelers. 
Doing so risks reinforcing their biases and subjectivity as they label the additional data. 
This narrow focus on a limited group of labelers and a specific labeling criterion restricts the model's ability to detect different facets of hate speech. 
For example, if the dataset primarily focuses on sexism, the trained model may struggle to identify hate speech targeting immigrants or other marginalized groups. 
In contrast, training the model with multiple datasets exposes it to a larger pool of labelers and a diverse range of hate speech facets. Each dataset brings a different perspective and definition of hate speech, encompassing various societal biases. 
The model learns from this wider range of interpretations, effectively reducing the impact of individual biases and enabling a more comprehensive understanding of hate speech. 

In summary, leveraging multiple smaller, unrelated data sets to learn jointly and transfer information between apparently unrelated learning tasks can help with performance, bias and subjectivity issues.
Future work will explore leveraging a more significant number of small datasets and analyze the impact on prediction subjectivity.

\section*{Declarations}
\begin{itemize}
	\item \textbf{Conflict of interest}: On behalf of all authors, the corresponding author states that there is no conflict of interest.
	\item \textbf{Data Availability Statement}: The datasets used in this study are available from the corresponding author on reasonable request.
	\item \textbf{Acknowledgements.} This research was supported by an Australian Government Research Training Program (RTP) Scholarship, by the Commonwealth of Australia (represented by the Defence Science and Technology Group) through a Defence Science Partnerships Agreement, and by the Australian Department of Home Affairs.
	This research was undertaken with the assistance of resources and services from the National Computational Infrastructure (NCI), which is supported by the Australian Government.
\end{itemize}

\section*{Acknowledgements}
This research was supported by an Australian Government Research Training Program (RTP) Scholarship, and from the Australian Department of Home Affairs. This research was undertaken with the assistance of resources and services from the National Computational Infrastructure (NCI), which is supported by the Australian Government.

\bibliographystyle{apa}

\clearpage
\appendix

\noindent This document is accompanying the submission \textit{\titlename}.
The information in this document complements the submission, and it is presented here for completeness reasons.
It is not required for understanding the main paper, nor for reproducing the results.

\section*{Supplemental figures}

\noindent This section presents the supplemental figures mentioned in the main text.
\begin{itemize}
	\item \cref{subfig:T2-conf-matrix-baseline,subfig:T2-conf-matrix-deephate} show the confusion matrixes made by Davidson baseline and \tdnn on the \dav data set;
	\item \cref{subfig:T6-barplots-davidson} Prediction performances with limited amounts of training data on the \dav dataset;
	\item \cref{si-subfig:hidden_size} shows the results of the grid search for the optimal size of the hidden state vector;
	\item \cref{si-subfig:batch_size} shows the results of the grid search for the optimal learning batch size;
	\item \cref{subfig:T4-MoH-general,subfig:T4-MoH-contextualized} depict the Map of Hate constructed by \dnn using general embeddings and task-specific embeddings, respectively.
	Visibly, when using general-purpose embeddings, the tweets belonging to different classes do not appear distinguishably separated, whereas they appear clustered when using task-specific embeddings.
	This highlights the impact of performing domain adaptation for textual embedding;
\end{itemize}

\begin{figure*}[t]
	\centering
	\newcommand\myheight{0.19} %
	\subfloat[]{
		\includegraphics[height=\myheight\textheight]{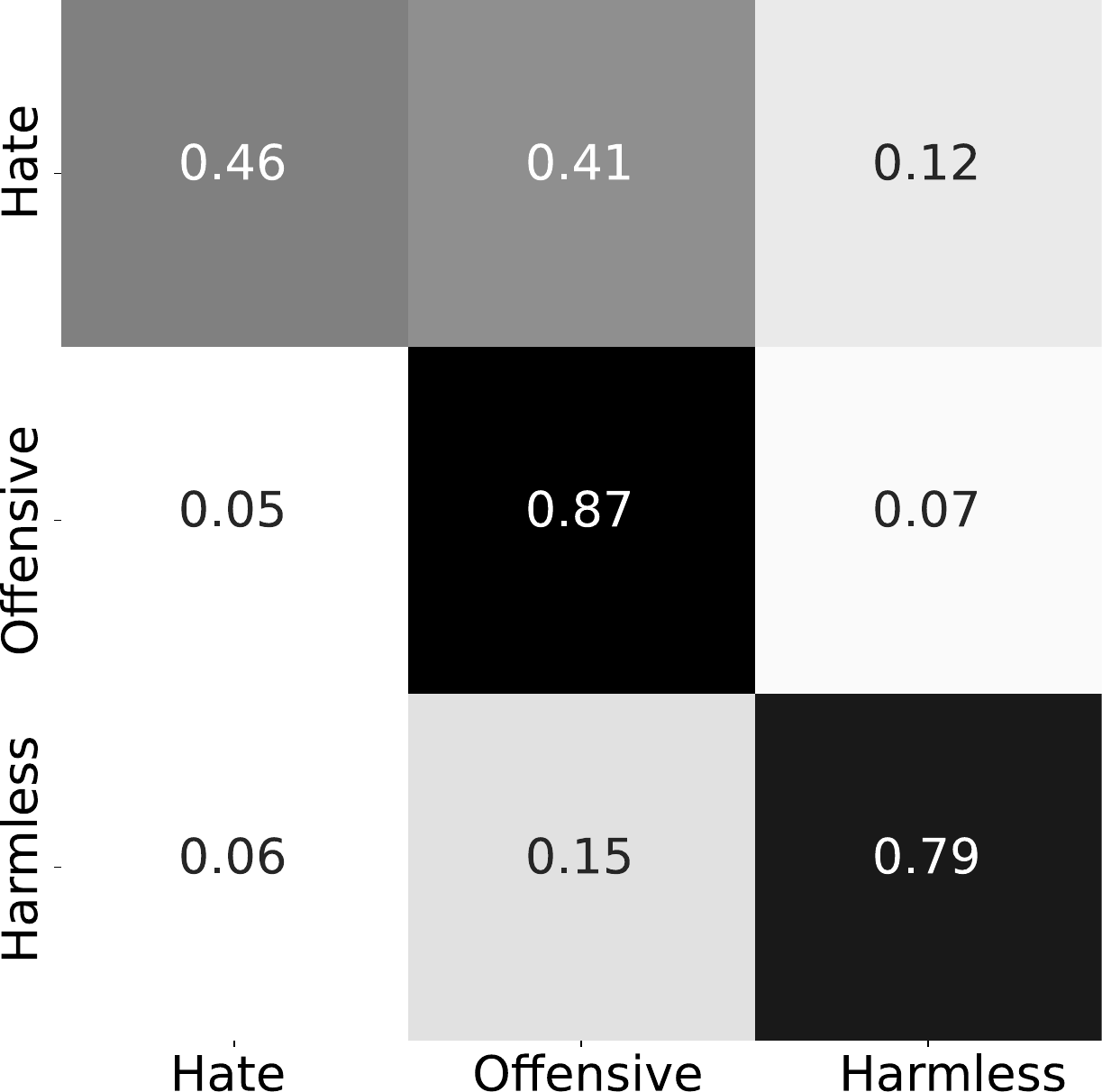}%
		\label{subfig:T2-conf-matrix-baseline}%
	}%
	\subfloat[]{
		\includegraphics[height=\myheight\textheight]{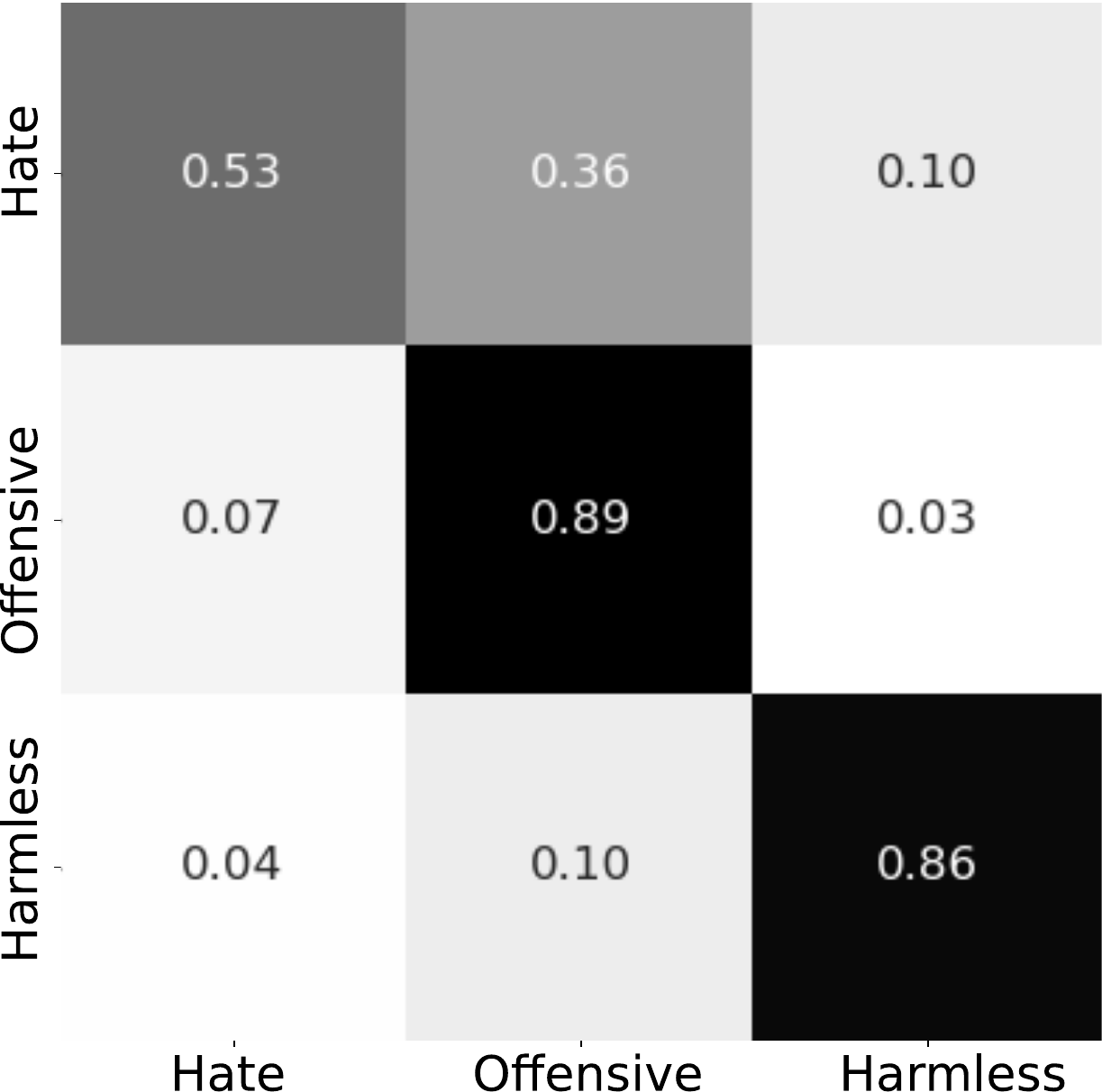}%
		\label{subfig:T2-conf-matrix-deephate}%
	}%
\\
	\subfloat[]{
		\includegraphics[height=\myheight\textheight,page=1]{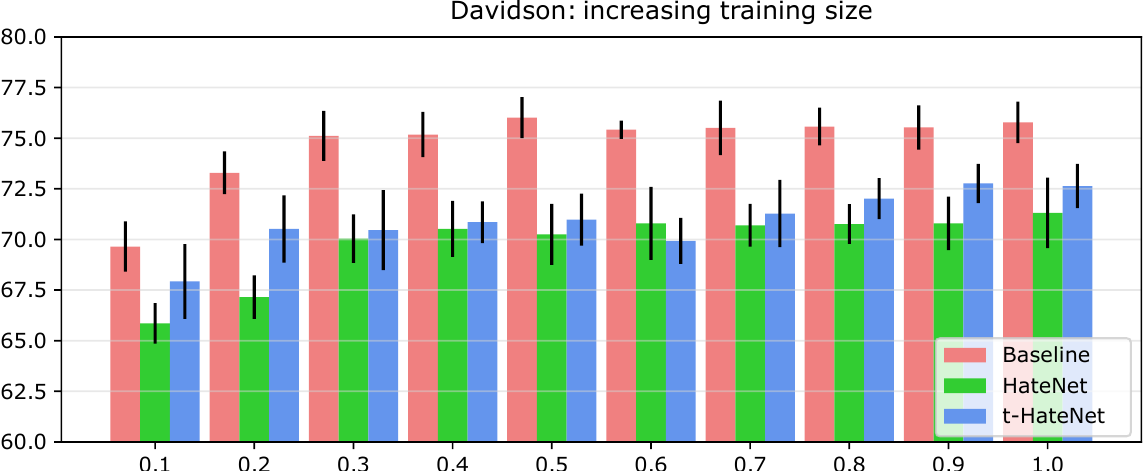}%
		\label{subfig:T6-barplots-davidson}%
	}

	\caption{
		\textbf{(a)(b)} Confusion matrix for one prediction made by the baseline (a) and by \tdnn (b) on the \dav data set.
      	\textbf{(c)} Prediction performances with limited amounts of training data on the \dav.
      	The x-axis show the percentage of the training set used for training, the y-axis shows the macro-F1 measure.
      	Each bar shows the mean value over 10 runs, and the standard deviation.
      }
      	
	\label{fig:T6}
\end{figure*} %
\begin{figure*}[htbp]
	\centering
	\newcommand\mywidth{0.48}
	\subfloat[]{
		\includegraphics[width=\mywidth\textwidth]{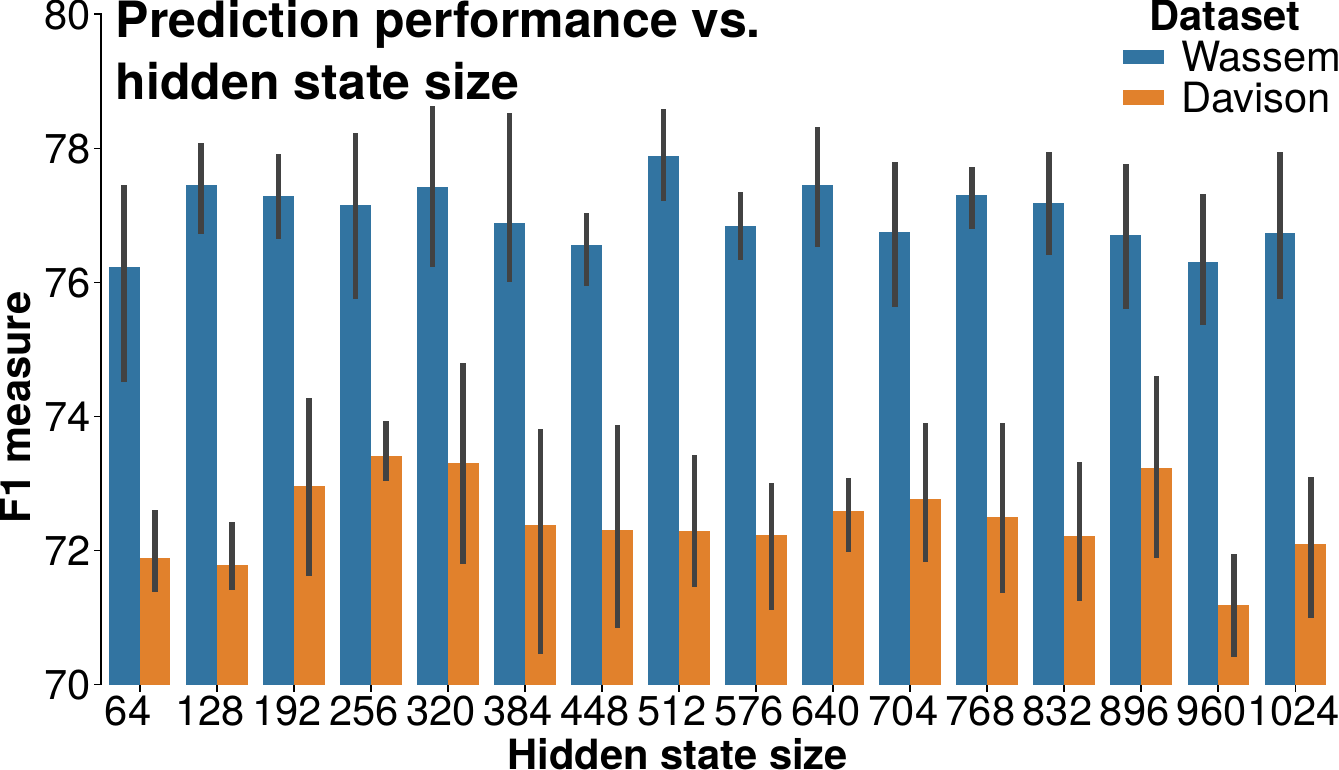}%
		\label{si-subfig:hidden_size}%
	}%
		\subfloat[]{
		\includegraphics[width=\mywidth\textwidth]{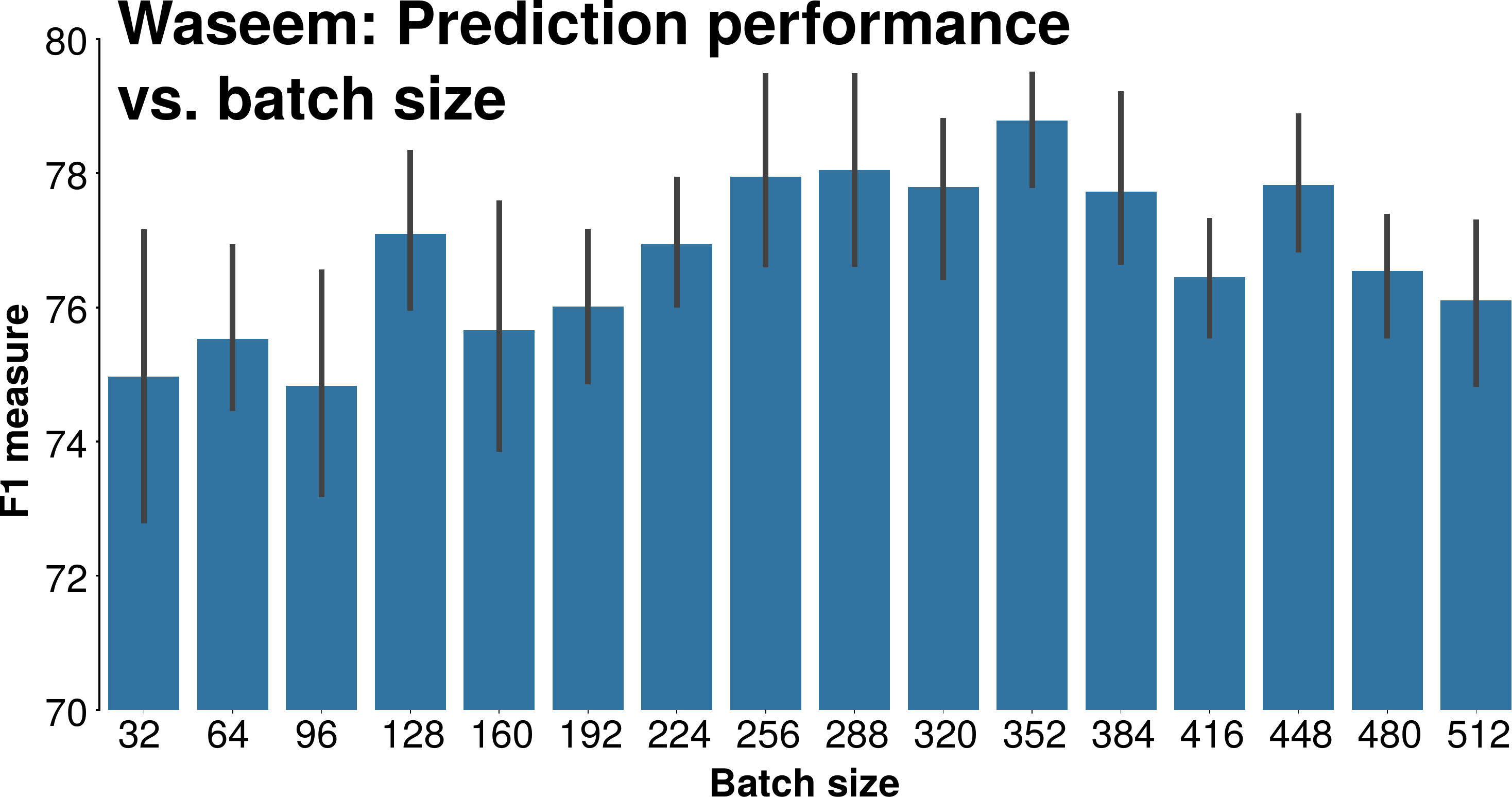}%
		\label{si-subfig:batch_size}%
	}%
	
	\caption{
      	Grid search for optimal hyper-parameter values: the hidden state size of the bi-LSTM (a) and the learning batch size (b).
	}
	\label{si-fig:T8}
\end{figure*} %
\begin{figure*}[htbp]
	\centering
	\newcommand\myheight{0.23}
	\subfloat[]{
		\includegraphics[height=\myheight\textheight]{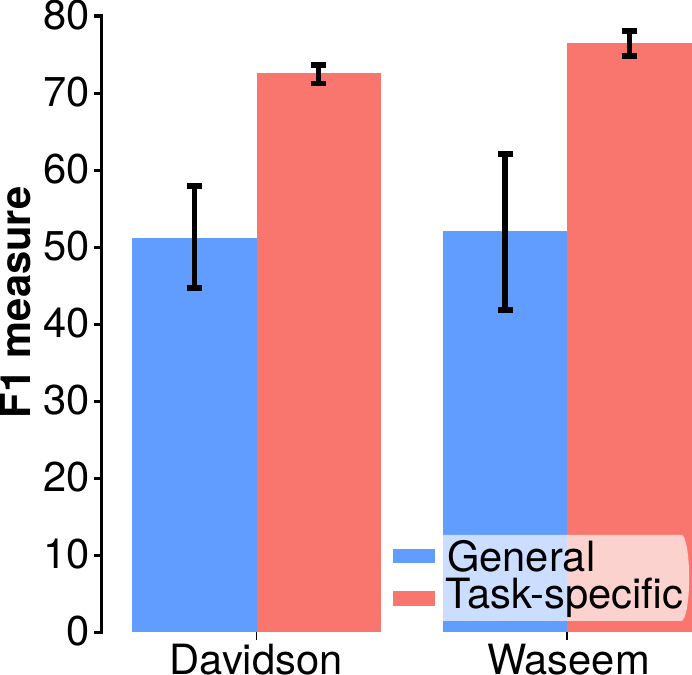}%
		\label{subfig:T4-prediction-performance-supp}%
	}\\%
		\subfloat[]{
		\includegraphics[height=\myheight\textheight,page=2]{T4_w_wo_LSTM}%
		\label{subfig:T4-MoH-general}%
	}%
	\subfloat[]{
		\includegraphics[height=\myheight\textheight]{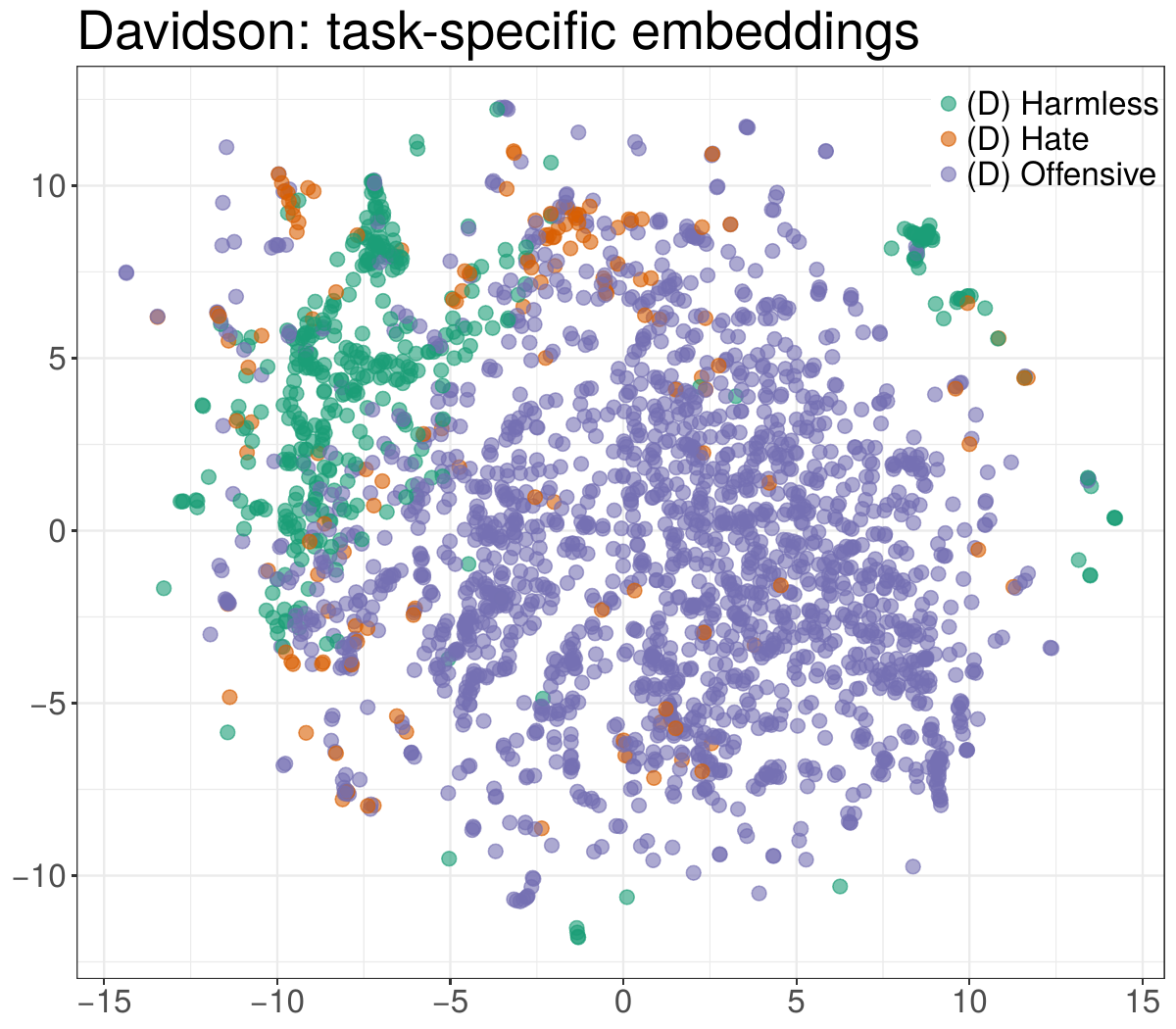}%
		\label{subfig:T4-MoH-contextualized}%
	}%
	
	\caption{
      	\textbf{The impact of building task-specific word embeddings.}
      	\textbf{(a)} The prediction performance (mean macro-F1 and standard deviation) of \dnn with general-purpose, and with task-specific embeddings.
      	\textbf{(b)(c)} The Map of Hate constructed using general-purpose embeddings (b) and using task-specific embeddings (c).
	}
	\label{fig:T4}
\end{figure*} 
 
\end{document}